\documentclass[conference]{IEEEtran}
\IEEEoverridecommandlockouts
\usepackage{cite}
\usepackage{amsmath,amssymb,amsfonts}
\usepackage{algorithmic}
\usepackage{graphicx}
\usepackage{textcomp}
\usepackage{xcolor}
\usepackage{booktabs}
\def\BibTeX{{\rm B\kern-.05em{\sc i\kern-.025em b}\kern-.08em
    T\kern-.1667em\lower.7ex\hbox{E}\kern-.125emX}}
\begin{document}

\title{Impact of network delays on Hyperledger Fabric}

\author{\IEEEauthorblockN{Thanh Son Lam Nguyen}
\IEEEauthorblockA{\textit{LIP6} \\
\textit{Sorbonne University}\\
Paris, France \\
Thanh-Son-Lam.Nguyen@lip6.fr}
\and
\IEEEauthorblockN{Guillaume Jourjon}
\IEEEauthorblockA{\textit{CSIRO-Data61} \\
Sydney, Australia \\
guillaume.jourjon@csiro.au}
\and
\IEEEauthorblockN{Maria Potop-Butucaru}
\IEEEauthorblockA{\textit{LIP6} \\
\textit{Sorbonne University}\\
Paris, France \\
maria.potop-butucaru@lip6.fr}
\and
\IEEEauthorblockN{Kim Loan Thai}
\IEEEauthorblockA{\textit{LIP6} \\
\textit{Sorbonne University}\\
Paris, France \\
kim.thai@lip6.fr}
}

\maketitle

\begin{abstract}
Blockchain has become one of the most attractive technologies for applications, with a large range of deployments such as production, economy, or banking.
Under the hood, Blockchain technology is a type of distributed database that supports untrusted parties. In this paper we focus
Hyperledger Fabric,  the first blockchain in the market tailored for a private environment, allowing businesses to create a permissioned network. Hyperledger Fabric implements a PBFT consensus in order to maintain a non forking blockchain at the application level.  
We deployed this framework over an area network between France and Germany in order to evaluate its performance when potentially large network delays are observed. 
Overall we found that when network delay increases significantly (i.e. up to 3.5 seconds at network layer between two clouds), 
we observed that the blocks added to our blockchain had up to 134 seconds offset after 100\textsuperscript{th} block from one cloud to another. 
Thus by delaying block propagation, we demonstrated that Hyperledger Fabric does not provide sufficient consistency guaranties to be deployed in critical environments. Our work, is the fist to evidence  the  negative impact of network delays on a PBFT-based blockchain. 
\end{abstract}

\section{Introduction}
Blockchains are distributed databases with no central authority and no point of trust. Unlike centralized systems or client/server databases where data is stored only in one or a group of servers, Blockchain database, also called ``ledger'', is stored on every peer in the network. Although being invented more than 30 years ago~\cite{b1}, blockchains received attention only after the explosion of Bitcoin~\cite{b2}. 

Bitcoin~\cite{b2}, Ethereum~\cite{b19} and many other popular blockchains are established on a public permissionless blockchain technology, opened to anyone, where participants interact anonymously. That is, in public or permissionless blockchains, everyone can join and use its services without providing any identity. 
Private or permissioned blockchains, in contrary, are  closed blockchains  open only to a set of known, identified users. 
Hyperledger Fabric~\cite{b5, b13, b20} is an 
  innovative project started at Linux Foundation~\cite{b9},  now led and managed by two  companies: IBM and Digital Asset. Hyperledger Fabric aims at providing a resilient, flexible,  and confidential blockchain framework. It is considered the foundation of private, open-source blockchain applied to business. Along the spree of Hyperledger Fabric development and usage since the first release in 2016, Fabric is being constantly improved into newest versions (version 1.4 at the end of 2018).
  
Hyperledger Fabric follows an  Execute-Order-Validate philosophy  while most of the existing PBFT blockchains implement an Order-Execute architecture.


\begin{figure*}
\centering
\includegraphics[width=290pt]{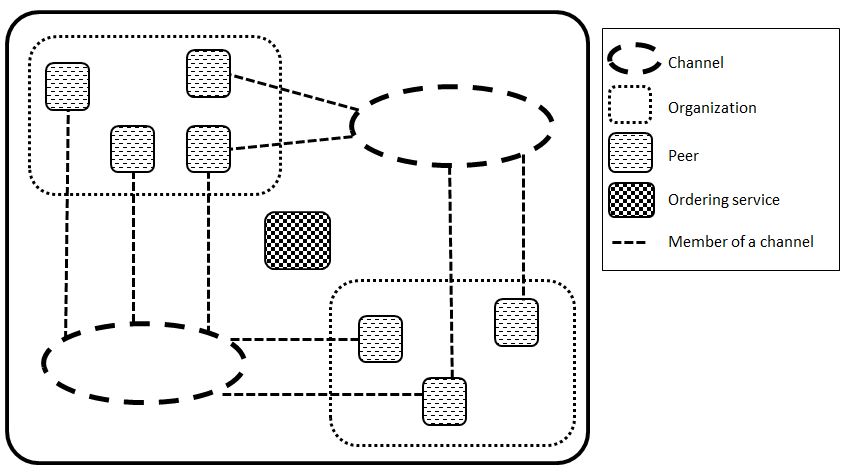}
\caption{Structure of Fabric: the main components.}
\label{fig2}
\end{figure*}

This  paper focus on  understanding the function of Hyperledger Fabric through the 3-phase procedure execute-order-validate. 
To do so we present a thorough explanation of how the components of Fabric interact with each other. Then we dive deep into the roles of Hyperledger Fabric at application and network layer. Then we evaluate its performances by varying delays at the network layer.  
We experiment a simple scenario where blocks are produced every 900ms. We increase the transmission delays  up to 3.5 seconds. This yield to dramatic blockchain desynchronisation. The 100\textsuperscript{th} block may be received by a peer with 134 seconds delay, making Hyperledger Fabric  not suitable for critical  applications such as banking or trading.

This paper is organized as follows: the next section briefly presents related work, Section~\ref{sec:fabric} discusses the main components of Hyperledger Fabric, while Section~\ref{sec:architecture} explains in details its Execute-Order-Validate architecture and the strength of this novel architecture with respect to the Order-Execute architectures. Section~\ref{sec:experiments} presents our methodology and the experimental results obtained with Hyperledger Fabric deployed on two clouds situated in France and Germany. Finally, Section~\ref{sec:conclusion} concludes our article.

\section{Related Work}
\label{sec:related}
There are two major trends in the design of blockchain systems: proof-of-work based blockchains (e.g. Bitcoin, Ethereum) and PBFT-based blockchains (e.g. Hyperledger, Tendermint, Byzcoin, or Bitcon-NG). 
In the Proof-of-Work (PoW) based blockchains, a miner must solve a complex puzzle  in order to have the ability to add a block to the blockchain. This is what the cryptocurrency world calls ``mining'' where clients ``prove'' that they have done the ``work''. This is a public blockchain which is based on old Order-Execute architecture and a perfect case study for the possible inefficiencies of blockchain that system takes about ten minutes to add a new block to the blockchain.  
Bitcoin is the most successful cryptocurrency in the last few years. Everybody can join into this blockchain network. 

One of the most representative PoW blockchains are Bitcoin and Ethereum(\cite{b3},\cite{b19}). Ethereum is 
the first blockchain that use smart contracts. Ethereum is similar to Bitcoin, as it offers a cryptocurrency but extends Bitcoin capability with smart contract.  
The theoretical study of proof-of-work based blockchains started with the analysis of the Bitcoin  agreement aspects in various synchronous models. The major criticisms for the proof-of-work approach are as follows: it is assumed that the honest miners hold a majority of the computational power, the generation of a block is energetically costly which yielded to the creation of mining pools and finally, the multiple blockchains that may coexist in the system.

In order to overcome these drawbacks, permissioned blockchains such as Tendermint~\cite{b4}, Byzcoin~\cite{b27}, or Bitcon-NG~\cite{b28}  have been proposed. These blockchains use Byzantine-fault tolerant (BFT) consensus \cite{b23} of Practical Byzantine-Fault Tolerant~\cite{b24} or other variants of these two classes of consensus protocols. However, they all follow the same Order-Execute approach. The limitations of Order-Execute architecture will be explained deeply in Section~\ref{sec:architecture}. Hyperledger Fabric  (\cite{b13}, \cite{b20}) builds on top of a PBFT protocol but its architecture follows a Execute-Order-Validate architecture that will be explained in details in the next section.

Blockchain systems, beyond their incontestable features such as decentralization, simple design and relative easy use, are not free of incidents and limitations. The most popular incident reported for Ethereum, for example, was the 60 million dollars Ether theft, which was possible by simply exploiting an error in the code and the lack of system specification.
Moreover, Blockchains are now the new attack targets of hackers around the world and also the main subject for the security researcher. One type of attack they can do is the hijacking attack, a type of network security attack in which the attacker takes control of the communication at the networking level.

In~\cite{b25} the authors describe two methods for hijacking Bitcoin: one by BGP hijacking and the second one by delaying propagation. They demonstrated that ``any network attacker can hijack few ($<$100) BGP prefixes to isolate $\sim$50\% of the mining power'', ``slow down block propagation by interfering with few key Bitcoin messages''.
 
Hijacking attack is also the premise for a full-fledged attacks on Ethereum blockchain to steal coins~\cite{b25}. They can multiply an asset by 200,000 in just 10 hours in consortium or private context.

To the best of our knowledge no previous work has been done in hijacking PBFT-based blockchains. 
In this work we target hijacking 
Hyperledger Fabric by delaying block propagation. Interestingly, this type of attack can be easilly performed although Hyperledger Fabric is a private blockchain where participants use their credentials and all the traffic between them is signed and encrypted.

\section{Fabric Components and Actors}
\label{sec:fabric}
In this section, we describe some important components of Hyperledger Fabric in more details. \figurename~\ref{fig2} shows some of them.

\textit{Smart contract or chaincode}- A smart contract, or ``chaincode'' called by Hyperledger Fabric, is a business logic or rules, represented by an application and invoked by a client application external to the blockchain network in order to manage access and modification to a set of key-value pairs in the ``World State'' of Ledger. It is installed onto peer nodes and started (instantiated) on channels.

\begin{figure}
\centering
\includegraphics[width=120pt]{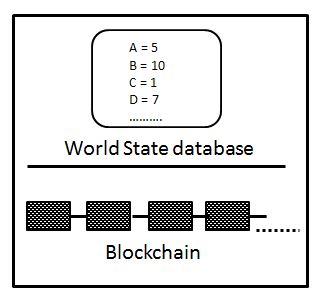}
\caption{Ledger: ``World State'' and ``blockchain''.}
\label{fig3}
\end{figure}

\textit{Ledger}- A ledger (see \figurename~\ref{fig3}) is a place to store the databases. There is only one logical ledger for each channel and every peer in that channel has the same copy of that logical ledger. It consists of two distinct sub-databases, though related, a ``blockchain'' and the ``world state'', also called as ``state database''. 

\begin{itemize}
\item ``Blockchain'' is a chain of blocks that each block contains a hash of the previous block and its data. These blocks are immutable, that means if once a block has been added to the chain, it cannot be changed or deleted.

\item In contrast, the ``world state'', also known as the ``current state'' or ``state database'', is a database containing the latest values for all keys in order to read, query, update or delete from chaincode efficiently. Peers commit the latest values to the ledger world state for each valid transaction included in a processed block. Supported databases include LevelDB~\cite{b8} and CouchDB~\cite{b7}.
\end{itemize}
\textit{Endorsement policy}- determines which endorsing peers on a channel must execute proposals and the required combination of responses. A validated transaction must satisfy the endorsement policy in order not to be marked as invalid by committing peers. This endorsement policy must be defined at the time the chaincode is started (instantiated).

\begin{figure}
\centering
\includegraphics[width=120pt]{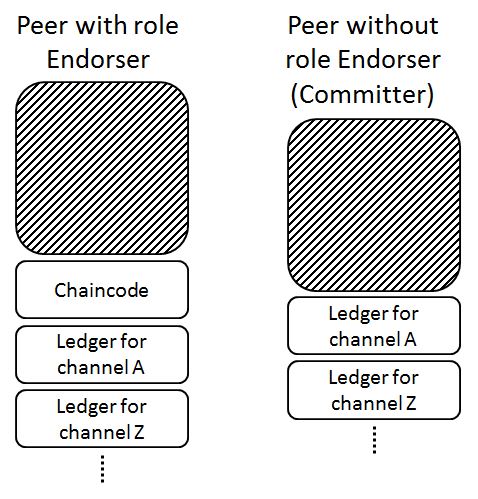}
\caption{Peer with the role Endorser and Committer.}
\label{fig4}
\end{figure}

\textit{Peer}- A blockchain network component, is owned and maintained by members, contains ledgers and may has chaincodes installed on in order to execute proposals from clients. \figurename~\ref{fig4} shows two main roles of the peer:
\begin{itemize}
\item Endorsing peer (Endorser): peer on which a chaincode was installed. It has the ability to execute and reply to the proposal attached to that chaincode. It is not mandatory that all peers in a channel are endorsing peers.

\item Committing peer (Committer): all peers in a channel are committing peers who will verify all transactions submitted before committing to and updating the ledger.
\end{itemize}

\textit{Ordering Service}- A collection of special nodes in the blockchain network called Orderers have a mission in ordering transactions into a block for each channel. Orderers maintain no ledger. There are two type of pluggable ordering service that are ``Solo'' and ``Kafka''. Solo is used for developing or testing purpose with only one Orderer node. Meanwhile, ``Kafka'' is intended for production with multiple Orderer nodes combine with Kafka/Zookeeper cluster~\cite{b11}. Following the documentation of Fabric~\cite{b13}, in the currently available releases, ordering service use Kafka and Zookeeper. In future releases, it will use a Raft~\cite{b6}\cite{b21} consensus ordering service instead.

\textit{Membership Service Provider (MSP)}- an abstract component that provides credentials to clients and peers. These credentials are used for clients to authenticate their transactions, meanwhile peers use to authenticate their proposal responses.

\textit{Proposal}- A request from client to endorsing peer (endorser) for endorsement. There are two types of proposal, one is Instantiate, another one is Invoke (read/write).

\textit{Transaction}- By gathering proposal responses from endorsing peers, client package the results and endorsements into a transaction for sending to Orderer. Invoke transaction performs read or write data operations from the ledger. Whilst, Instantiate transaction  starts and initializes a chaincode on a channel.

\textit{Channel}- an abstract relation between organizations. Each channel keeps data isolated by having different ledger which is shared across the peers in the same channel. Channel is defined by the Genesis Block (the first block of the blockchain on the ledger) and Configuration Block.

\textit{Organization}- the entities that own the peers. By adding its Membership Service Provider (MSP) to the network, an organization is joined to that network, map to more than one MSP. Using MSP and valid identity issued by organizations, members of the network may verify signatures each other (e.g. over transactions).
 
\section{Hyperledger Fabric Architecture}
\label{sec:architecture}

In this section, we describe briefly  Execute-Order-Validate paradigm in Hyperledger Fabric and explain the limitations of Order-Execute paradigm.

\subsection{Order-Execute Architecture}
Most of the blockchain systems precedent to Hyperledger Fabric used an Order-Execute architecture. Following this architecture, first, it orders the transactions and propagates to all the peers. Second, each peer executes the transactions sequentially. This paradigm is illustrated by \figurename~\ref{fig5} below.

\begin{figure}[!hbt]
\centering
\includegraphics[width=150pt]{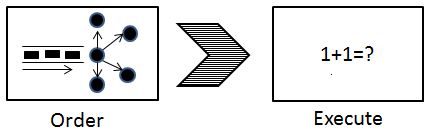}
\caption{Order-Execute architecture.}
\label{fig5}
\end{figure}

Although this architecture is simple and widely used, it has several limitation when apply onto permissioned blockchain. Here are some important limitations:
\begin{itemize}
\item Transactions must be deterministic. The transactions are ordered first and then sent to all peers to execute. In this second step, if the transactions are not deterministic, there is no guarantee  of the same result after executing and the ledger can be different from peer to peer. 
\item All the peers must maintain the smart contract to execute the transactions, lead to transaction data or smart contract logic be easily to view. This case violates the confidentiality which is required in permissioned blockchain.
\item The sequential execution of all transactions by all peers limits the performance, lead to bottleneck and easy denial-of-service (DoS) attacks.
\end{itemize}

\subsection{Execute-Order-Validate architecture of Fabric}
\begin{figure}[!hbt]
\centering
\includegraphics[width=\linewidth]{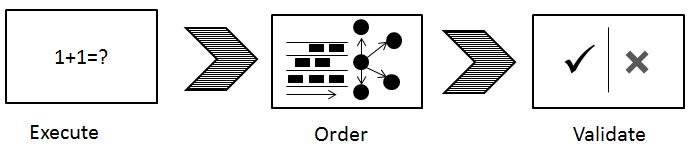}
\caption{Three phases architecture of Hyperledger Fabric.}
\label{fig6}
\end{figure}

Due to some limitations of the paradigm above, Hyperledger Fabric adopts  Execute-Order-Validate paradigm (showed in \figurename~\ref{fig6}). This new architecture is resilient, flexible, scalable, confident with modular design for permissioned blockchains.

\paragraph{Execution phase} Clients sign and send the proposal to the endorsers and wait for their reply. The endorsers, the peers who have the chaincode installed, execute the operation on the chaincode individually without synchronization with other endorser peers. Endorsers  create two values called Writeset and Readset with the result of the operation and then cryptographically sign this endorsement, send it back to the client in a proposal response. The client that receives these responses verify if they satisfy the endorsement policy then it  creates a transactions that assembles these responses for the next phase, Ordering phase.

\paragraph{Ordering phase} Client sends a transaction  to the ordering service after receiving enough proposal responses from endorser peers. The ordering service order all submitted transactions per channel. Moreover, in order to improve the throughput, the ordering service combines transactions for a channel into blocks. Ordering service, then, broadcasts blocks to all the peers in a channel.

\paragraph{Validating phase} Each peer in a channel after receiving blocks from ordering service will run the following three sequential steps:

\begin{itemize}
\item Verify all transactions within the block that must satisfy the endorsement policy. If the endorsement is not satisfied, the transaction is marked as invalid.

\item Check all transactions in the block sequentially by comparing the versions of the key in the READSET field to those in the World State. If the versions do not match, the transaction is marked as invalid.

\item Append the block to the blockchain and update the ledger as well.
\end{itemize}
The \figurename~\ref{fig7} below gives an overview of the whole picture how a block is added into the blockchain in Hyperledger Fabric by showing the process from proposal transmission to updating ledger. 

\begin{figure}[!hbt]
\centering
\includegraphics[width=\linewidth]{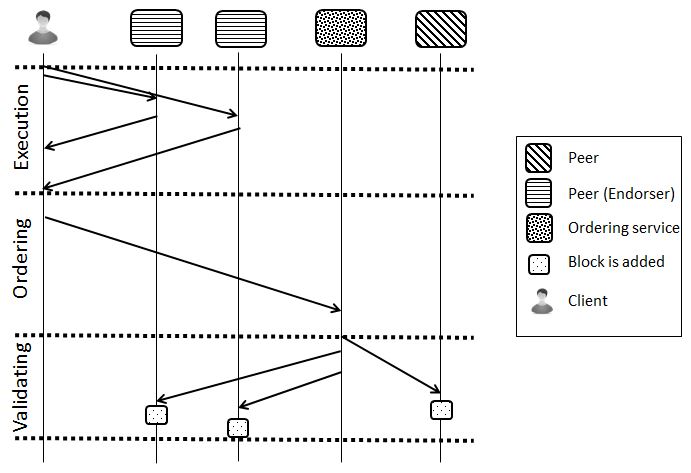}
\caption{High level transaction flow.}
\label{fig7}
\end{figure}

\section{Experiments}
\label{sec:experiments}
\begin{figure}[!hbt]
\centering
\includegraphics[width=200pt]{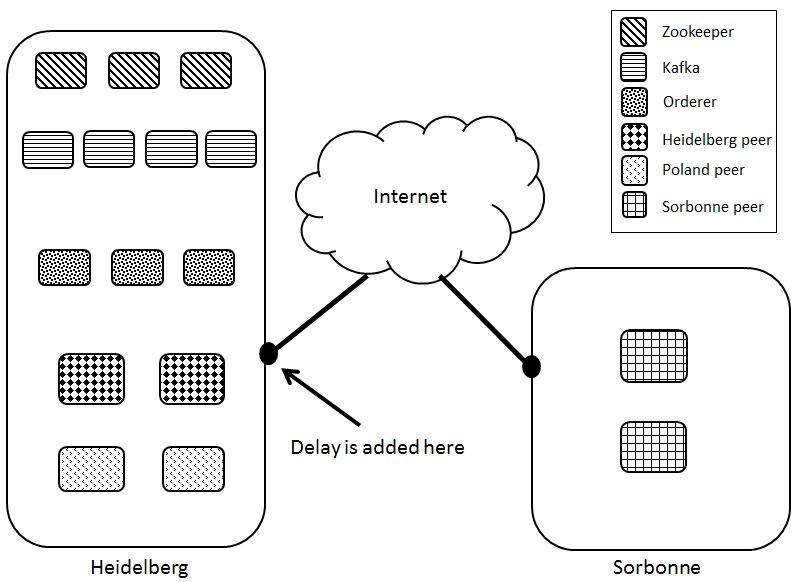}
\caption{Topology of this experiment.}
\label{fig8}
\end{figure}
This experiment will focus on the delay between nodes in Hyperledger Fabric and its impact on the update block into the ledger processes.

\subsection{Setup}
\begin{figure*}[!hbt]
\centering
\includegraphics[width=400pt]{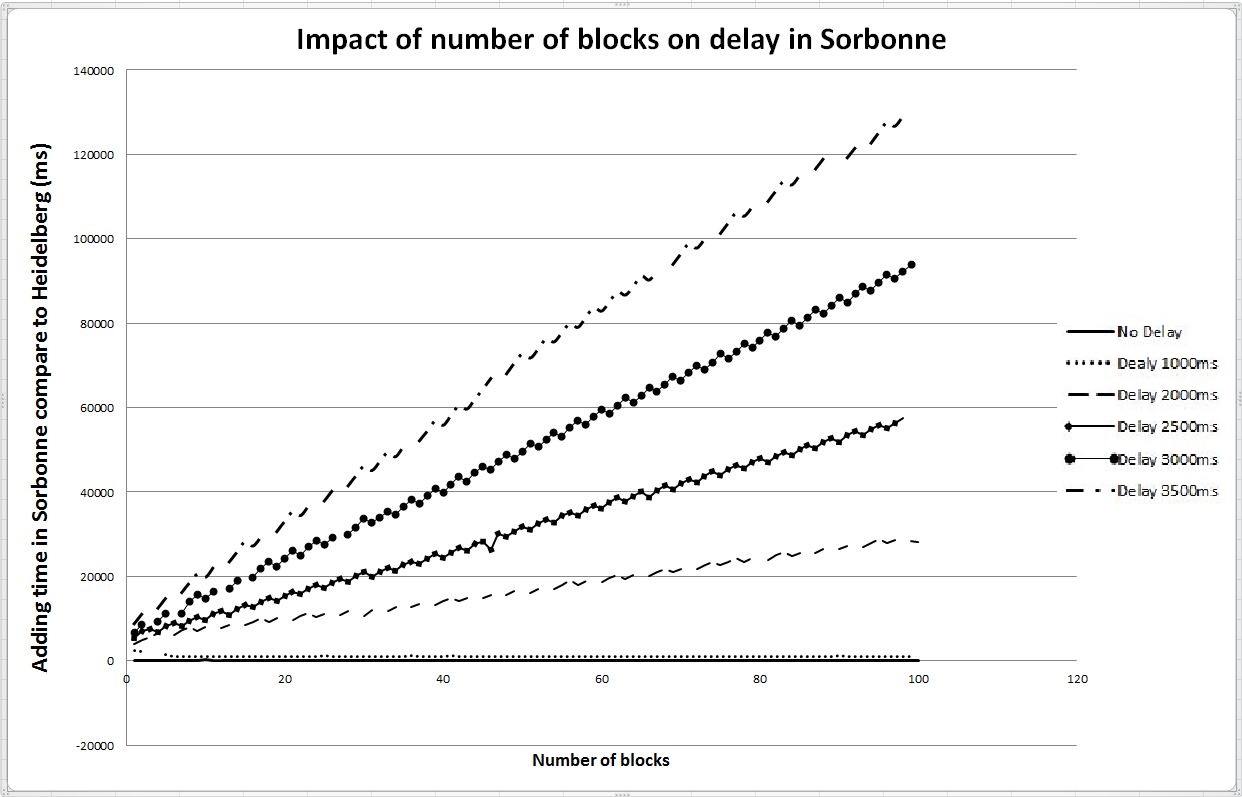}
\caption{The result of most important delays.}
\label{fig9}
\end{figure*}
In this experiment, nodes are hosted in 2 cloud instances: OneLab (https://onelab.eu/team,  LIP6, Sorbonne University) and the second from Heidelberg University. Each cloud instance has 2 vCPU (2.6GHz in Sorbonne and 2.4GHz in Heidelberg), 4GB of RAM, runs Ubuntu 16.04.5 LTS. Version of Docker [12] using is 18.09.0. On this infrastructure, a Hyperledger Fabric blockchain network in version 1.2.1 is created. There is a single channel on Kafka ordering service type with 3 Zookeepers nodes, 4 Kafka brokers and 3 orderers, all on distinct VMs. There are also 6 peers in total that belong to 3 organizations called: "Sorbonne", "Heidelberg" and "Poland". Each organization has 2 peers (peer0 and peer1), one of them is endorser (peer0). The communication uses TLS (Transport Layer Security). Moreover, Network Time Protocol (NTP) service is used to synchronize nodes clocks.

The tool to setup delays  is TC~\cite{b14} (Traffic Control) of Linux. 

\subsection{Methodology}

In every experiments, we use the same endorsement policy “AND ("Heidelberg" peer, "Poland" peer)” in order to satisfy that the Execute and Orderer phase will not affected to Sorbonne peer. "Sorbonne" peers only  update blocks into the ledger. The delay between the two clouds will be increased up to lost connection. Based on this, in each delay different, we flood 1000 transactions sequentially. We will observe the time a block is added in "Sorbonne" peers compared to the time that block will be added in Heidelberg. In this way we will know the time duration peers in Sorbonne need to wait compare to Heidelberg peer just for adding the same block into the blockchain.

\subsection{Results}
In our experiments we used Hyperledger Fabric version 1.2.1. The transactions were sent sequentially one after another about 85ms in average. Blocks were create each 900ms, 46 Kbytes in size, containing 10 transactions. Using the command PING to detect the duration between Heidelberg and Sorbonne cloud, we achieved 21.7ms in average.

We run the experiment with different delay values and for each delay value we execute 5 runs.  Table~\ref{tab1} indicates the offset time when add a block into ledger on Sorbonne compared to that block in Heidelberg. \footnote{All experiment result are stored on google drive \cite{b22}.}

\begin{table}\scriptsize
\caption{Offset values between Sorbonne and Heidelberg}
\begin{center}
\begin{tabular}{ccccccc}
\toprule
ith-Block&No delay&1000ms&2000ms&2500ms&3000ms&3500ms \\
\midrule
1&84.2&2288.6&4056&5566&6858&8778  \\
10&168&985.2&7893&9959&14782&19801  \\
19&18.8&972.4&10038&14312&22468&30605  \\
30&72&951.8&10593&21190&33642&46116  \\
40&90.8&1042.8&14239&24631&39864&55820  \\
49&37.8&1046.4&16392&30685&47866&70567  \\
60&45.2&1036&18553&36192&59530&82877  \\
70&92.4&1044.4&21825&42040&66445&96450  \\
79&74.6&1023.8&24248&47042&74199&107545  \\
88&58&1066&26382&51754&82357&119012  \\
97&95.6&974&28674&56469&90585&126680  \\
\bottomrule
\end{tabular}
\label{tab1}
\end{center}
\end{table}

In case of delays smaller than 2s, the offset of adding a block in Sorbonne and Heidelberg depend on the delay value. In general, the offset time between them somehow we can consider the same.

When delays are greater than or equal to 2s, this offset time increases with the number of blocks. For example, in case of delay 2s, when the first block is added in Heidelberg in time t0, that first block will be added later in t1 = t0 + 4056ms, just for 2s delay. And this offset is increased when more blocks are added, it takes more than 28 seconds for the 100\textsuperscript{th} block. If the delay is set to 3500ms, the offset is increased from 8778ms up to 128 seconds for the first blocks and 100\textsuperscript{th} block. In this situation, that means a query for an asset in Heidelberg will be up to date value, on the contrary, the same query will get an old value of that asset from the Sorbonne site in the same time.

Moreover, we run one experiment using 3500ms delay and we increased the number of transactions up to 30000, there were nearly 3000 blocks submitted and the offset now was increased up to 1 hour and 10 minutes.

When we increased the delay to 3580ms, we observed that after few blocks added in Sorbonne the system stopped. By verifying in the network layer,  the Sorbonne node was considered as disconnected to the docker swarm. That is the cause the Sorbonne node cannot receive any blocks from the blockchain network to update the ledger.

In order to understand  this unexpected disconnection we investigate the communications using Wireshark tool the following scenario. The communication delay was set to  3500ms. We oserved that each time a peer received a block, this peer sends back to Orderer a signal. When the Orderer received that signal, it sends the next block to the peer. With the delay set to 3500ms, the transmission time was increased by this delay and linearly increases with the number of blocks. 

Moreover, this offset will be increased due to the buffer size of the Orderer. If this buffer is big, the Orderer can store more blocks and the offset increases. When the buffer reaches its limit, the Orderer is considered as halting which affects the whole system.

\begin{figure}[!hbt]
\centering
\includegraphics[width=.9\linewidth]{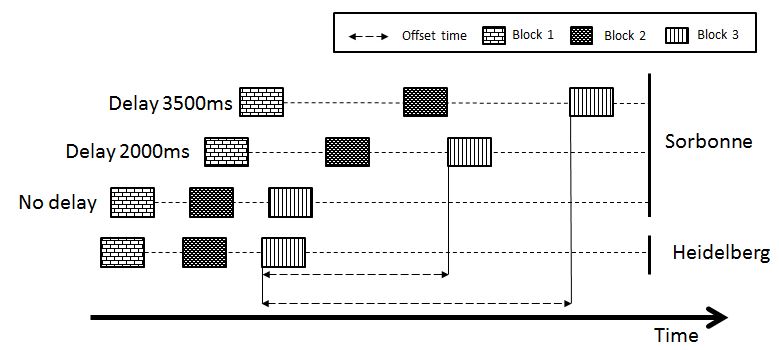}
\caption{Impact of small network delay  on the offset.}
\label{fig10}
\end{figure}

In consequence, small network delays have a tremendous impact on the offset. \figurename~\ref{fig10} illustrates this result.

\section{Conclusion}
\label{sec:conclusion}
This paper is the first to present the impact of network delays on a PBFT-based blockchain, i.e. Hyperledger Fabric. We deployed Hyperledger Fabric  on two clouds situated in France and Germany and introduced transmission delays up to 3.5 seconds. In our experiments we observed that  the 100\textsuperscript{th} block is updated in the two ledgers  with a delay  up to 134 seconds when the transmission delay is 3.5 seconds. Moreover, the system brutally halts when delays are greater than 3.5 seconds. We conclude that the tested version of Hyperledger Fabric (the most up to date at the time of our experiments) cannot be used in critical applications such as banking or trading. 
 
Our study extends the recent work on hijacking permissionless blockchains and advocates that existing blockchains architectures should be redesigned in order to be resilient to network attacks. 

As future work we intend to extend our study to other types of blockhains (Tendermint, IOTA, Hashgraphs) and to long distance connections. We currently plan replay our experiment by deploying Hyperledger Fabric on two clouds situated in France and Australia. 

\section{Acknowlegments}
The authors would like to thank prof. Vincent Heuveline, head of Interdisciplinary Center for Scientific Computing (IWR) at 
Engineering Mathematics and Computing Lab (EMCL), Heidelberg University and Olaf Pitcher for facilitating our access to the  Heidelberg University cloud.
 We thank prof. Pascal Frey the head of Institut des Sciences du Calcul et des Données (Sorbonne University) for deep discussions on the "Genomic on blockchain" joint project between Sorbonne University and Heidelberg University. The first author has been supporter by a grant offered by the Institut des Sciences du Calcul et des Données, Sorbonne University.

\end{document}